\documentclass[prb,twocolumn, superscriptaddress, showpacs, a4paper]{revtex4}

\usepackage{graphicx}

\newcommand{\ket}[1]{\ensuremath{\left|#1\right\rangle}}

\begin{document}

\title{III-nitride based quantum dots for photon emission\\with controlled polarization switching}

\author{S. Amloy}
\affiliation{Link\"oping University, Semiconductor Materials Division, \\Department of Physics, Chemistry, and Biology (IFM), SE-58183 Link\"oping, Sweden}
\affiliation{Thaksin University, Department of Physics, Faculty of Science, Phattalung 93110, Thailand}
\author{K. F. Karlsson}
\affiliation{Link\"oping University, Semiconductor Materials Division, \\Department of Physics, Chemistry, and Biology (IFM), SE-58183 Link\"oping, Sweden}
\email{freka@ifm.liu.se}
\author{P. O. Holtz}
\affiliation{Link\"oping University, Semiconductor Materials Division, \\Department of Physics, Chemistry, and Biology (IFM), SE-58183 Link\"oping, Sweden}

\date{\today}

\begin{abstract}
Computational studies based on 6 band k$\cdot$p theory are employed on lens-shaped III-nitride quantum dots (QDs) with focus on the polarization properties of the optical interband transitions. The results predict pronounced linear polarization of the ground-state related transitions for asymmetric QDs of a material with small split-off energy. It is demonstrated that a moderate externally applied electric field can be used to induce a linear polarization and to control its direction. InN is found to be the most efficient choice for dynamic polarization switching controlled by an electric field, with potential for polarization control on a photon-by-photon level.
\end{abstract}

\pacs{73.21.La, 73.22.Dj, 78.20.Bh, 78.67.Hc}

\maketitle

\section{Introduction}

A challenging field of applications for single quantum dots (QDs) belongs to quantum information technologies, including quantum cryptography and optical quantum computing. These interesting applications exploit single- and entangled-photons on demand. \cite{ref01, ref02, ref03} Single-photons on demand are typically obtained by pulsed excitation in combination with spectral filtering selecting a specific excitonic emission line of the QD. Control of the optical linear polarization can be archived by direct control of the dot symmetry at growth. \cite{ref04} For applications that require switching of the polarization on a photon-by-photon level, the dot symmetry can be modified dynamically by an externally applied uniaxial stress.\cite{ref05} Alternatively, an external electric field can be used to modify the valence band character and thereby control the polarization of the QD emission\cite{ref06} or to tune the nominally unpolarized QD emission into polarized resonances of an optical micro cavity.\cite{ref07}

The III-nitrides possess unique properties making them very attractive as the active material of a single photon source. The III-nitrides exhibit a wide range of possible band gaps, which can be exploited for spectral tunability from the infrared to deep ultra-violet frequencies. Moreover, the large band gap offsets provide deep carrier confinement allowing a high operating temperature.\cite{ref08, ref09} The nitrides can crystallize in either the wurtzite or the zinc-blende structure, but wurtzite is the most thermodynamically stable structure. However, the zinc-blende structure can be stabilized by epitaxial growth on cubic substrates. \cite{ref10, ref11, ref12, ref13} The wurtzite and the zinc-blende structures can have a large strain-induced piezoelectric polarization along the [0001] and [111] directions, respectively, which strongly affects the confinement potential that in turn may influence the polarization properties of the optical transition. 

In this work, it will be demonstrated that III-nitrides also are very suitable for attaining polarization control of the emitted photons. This study is based on numerical computations employed on lens-shaped III-nitride and III-arsenide QDs in order to elucidate the effects of the split-off energy, QD symmetry and external electric fields on the linear polarization properties of ground-state related interband transitions.  These computations based on our theoretical model highlight the III-nitrides as potentially important for the design of QD structures for control of the optical polarization of single photons, with various possible applications in the field of polarized light, such as polarization sensing and quantum information technologies. \cite{ref14}

\section{Theoretical approach}
The electron (hole) envelope wavefunctions and energy levels are obtained by means of a single band (6 band) k$\cdot$p Hamiltonian, discretized by finite differences. The full three-dimensional solutions are computed for the wurtzite and the zinc-blende k$\cdot$p Hamiltonians based on Refs. \onlinecite{ref15} and \onlinecite{ref16}, respectively, using material parameters\cite{note1} from Refs. \onlinecite{ref19} and \onlinecite{ref21}. The Hamiltonians include strain, as computed with continuum elastic theory, \cite{ref17} as well as internal electric fields originating from the spontaneous and piezoelectric polarizations. The optical transitions are computed from the dipole matrix elements and the polarization properties are determined in terms of the degree of polarization ($P$), \begin{equation}
 P = \frac{I_x-I_y}{I_x + I_y}
\end{equation}where $I_x$ and $I_y$ are the transition intensities polarized along two orthogonal directions, $x$ and $y$, in the base plane of the QDs. The in-plane directions are $x:[1\bar{1}00]$ and $y:[11\bar{2}0]$  for wurtzite QDs on the (0001)-plane,  $x:[1\bar{1}0]$ and $y:[110]$ for zinc-blende QDs on the (001)-plane and $x:[1\bar{1}0]$ and $y:[11\bar{2}]$ for zince-blende QDs on the (111)-plane. A QD structure of lens-like shape with fixed height $h$ = 2 nm and an elliptical base enclosing the area $\pi r^2$ with a mean radius of $r$ = 6 nm was used in this model. In order to investigate the effects of elongation, a QD symmetry parameter, $\alpha$, is defined as the ratio between the minor (b) and the major (a) axes of the elliptic base (see the inset of Fig. 1). The major axis is always in the $x$-direction.

\begin{figure}[top]
\includegraphics[width=8 cm]{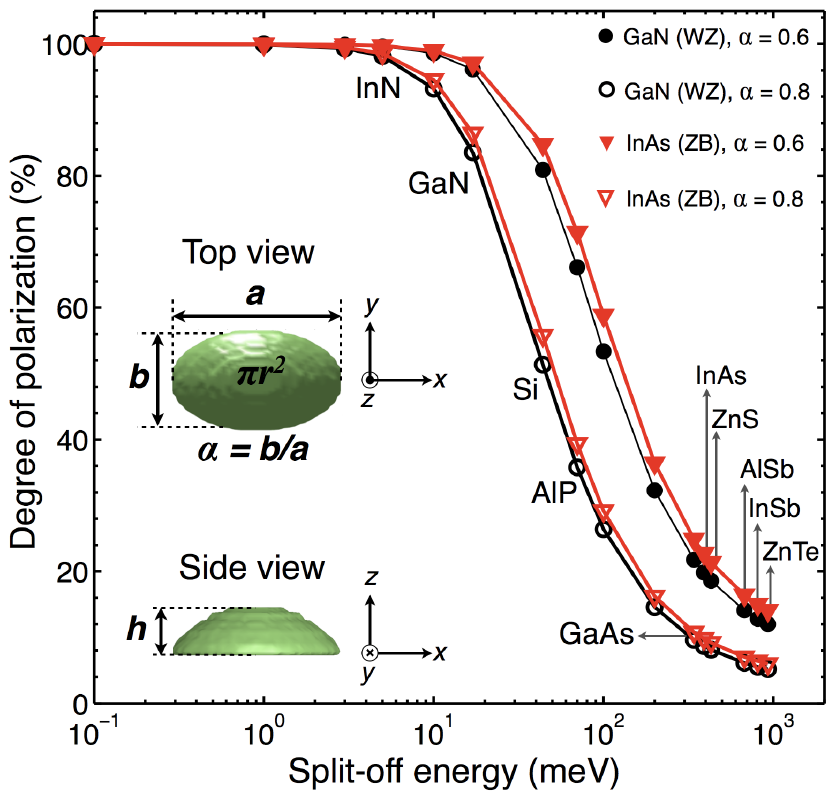}
\caption{
(Color online) The computed linear polarization degrees of the ground state related interband transition of lens-shaped QDs based on wurtzite (circles) and zinc-blend (triangles) Hamiltonians using the actual material parameters of GaN and InAs, except for the split-off energy which is varied. Two different symmetry parameters 0.6 and 0.8 are indicated by filled and open symbols, respectively. The labels indicate the actual split-off energies of some semiconductors. Solid lines serve as a guide to the eye. The inset shows a schematic illustration of the QD model in top and side view, respectively. The coordinates $x$, $y$, and $z$ refer to the $[1\bar{1}00]$, $[11\bar{2}0]$, and $[0001]$  ($[1\bar{1}0]$, $[110]$, and $[001]$) crystallographic directions in a wurtzite (zinc-blende) structure, respectively. 
}
\end{figure}

\section{The polarization dependence on the split-off energy}

The Bloch function of the conduction band inherits its properties from the atomic $s$-orbitals, while the top of the valence band is associated with three Bloch functions $\ket{X}$, $\ket{Y}$, and $\ket{Z}$, corresponding to directed $p$-orbitals ($p_x$, $p_y$ and $p_z$). For bulk or quantum wells without spin-orbit interaction, $\ket{X}$, $\ket{Y}$, $\ket{Z}$ are eigenstates of the valence band at the Brillouin zone center, implying that the optical transitions between the conduction band and the valence band can be purely linearly polarized in the $x$-, $y$-, or $z$-directions. The spin-orbit interactions couple $\ket{X}$, $\ket{Y}$, and $\ket{Z}$ into new eigenstates given by a superposition, e.g. for a conventional zinc-blende quantum well, the top of the valence band is given by $\frac{1}{\sqrt{2}}( \ket{X \uparrow} + i \ket{Y \uparrow})$ and $\frac{1}{\sqrt{2}}( \ket{X \downarrow} - i \ket{Y \downarrow})$, corresponding to degenerate heavy holes with spin up ($\uparrow$) or down ($\downarrow$). The corresponding optical transitions, completely polarized in the $x-y$ plane, do not exhibit any in-plane polarization anisotropy. If the symmetry of the quantum well is \emph{weakly} broken, e.g. by uniaxial lateral stress, these heavy-hole states will remain \emph{approximate} eigenstates and the optical transitions will still be essentially polarization isotropic. This is in contrast to the case without spin orbit interaction, where a weak symmetry breaking lifts a single directed $p$-orbital to the top of the valence band, consequently with perfectly linearly polarized optical transitions. Asymmetry can also be caused by anisotropic lateral quantum confinement, e.g. provided by QDs, but in this case are the valence band eigenstates not of pure $\ket{X}$, $\ket{Y}$, or $\ket{Z}$ character. Thus, the band mixing caused by lateral confinement in QDs may lower the degree of polarization from unity.

A measure of the spin-orbit interaction is the split-off energy ($\Delta_{SO}$), and its value varies widely from semiconductor to semiconductor, from 5 meV for InN to 930 meV for ZnTe. It is therefore interesting to investigate how the optical polarization of asymmetric QDs depends on $\Delta_{SO}$. According to the arguments presented above, it is expected that, for a given asymmetry, the degree of polarization is higher for smaller values of $\Delta_{SO}$. 

The split-off  energy of  the  InN, GaN, and AlN materials are 5, 17, and 19 meV, respectively,\cite{ref19} to be compared with the corresponding values of 380, 341, and 280 meV for the arsenide materials InAs, GaAs, and AlAs, respectively.\cite{ref20, ref21} In order to study the polarization dependence on $\Delta_{SO}$, the computations were performed on QDs of two commonly studied material systems, i.e. wurtzite GaN (0001) and zinc-blende InAs (001). By letting $\Delta_{SO}$ range from 0.1 to 930 meV, the two models yield very similar trends of the polarization degree for transitions between electrons and holes in their ground states, nearly independent on the crystal structure, as shown in Fig.1. The resulting polarization degree varies as expected in a wide range, with highest values $\sim$100\% for small $\Delta_{SO}$ down to $\sim$12\% for $\Delta_{SO}$ = 930 meV. It is worth to note that the degree of polarization indeed approaches 100\%, when $\Delta_{SO}$ approaches zero, indicating that the mixing of directed orbitals in the QDs states is insignificant. For GaN with the actual value $\Delta_{SO}$ = 17 meV, the degree of polarization is higher than 95\%, while the corresponding degree of polarization for InAs ($\Delta_{SO}$ = 390 meV) is only $\sim$20\% for the symmetry parameter $\alpha$ = 0.6. For a lower in-plane anisotropy ($\alpha$ = 0.8) the degree of polarization is weaker than for the higher anisotropy ($\alpha$ = 0.6) exhibiting the expected progressive decrease of the polarization degree as the QD becomes more symmetric. These results clearly show that $\Delta_{SO}$ is a main parameter affecting the degree of polarization for anisotropic QDs. It should be noted that a very high degree of linear polarization has previously been theoretically predicted for III-nitride QDs,\cite{ref22} as well as explained in terms of the energy splitting of the top valence bands, \cite{ref23} but the direct relation between high degree of polarization and $\Delta_{SO}$ was never clarified.

\section{The polarization dependence on the QD symmetry}

The effects of in-plane anisotropy originating from QD elongation in wurtzite GaN and zinc-blend InAs on valence bands are investigated for the actual values of $\Delta_{SO}$. The envelope wavefunctions of the electron and the hole ground states are both $s$-like, but they are modified by the elongated potential, and they will align parallel to the major axis of the QD. Optical transitions between the hole and the electron ground states is to some degree linearly polarized along the major axis. As the symmetry parameter is varied from 1.0, corresponding to symmetric QD, down to 0.2, the transition energy is blue-shifted by $\sim$115 meV ($\sim$36 meV) for the GaN (InAs) QD, from the emission energy $\sim$3.757 eV ($\sim$1.305 eV) for $\alpha$ = 1. The corresponding degree of polarization induced by the dot anisotropy is shown in Fig. 2. The GaN QD reveals a large variation of the polarization degree, with a resulting decrease from $\sim$100 down to 0\%, whereas a considerably smaller variation of the polarization degree is observed for InAs QD (from $\sim$60 down to $\sim$0\%). It is also notable that the GaN QD exhibits high and nearly constant polarization degree over a larger range of the symmetry parameter than the InAs QD. The origin of these strikingly different dependencies mainly refers to the significantly smaller split-off energy ($\sim$23 times) for GaN/AlN QD compared to InAs/GaAs QD. The results are in agreement with typical reported experimental values $P<40\%$ for InAs QDs.\cite{ref04} In contrast, polarization degrees have been observed ranging from 30-90\% for GaN/Al(Ga)N QDs\cite{ref23, ref25, ref26} and 20-96\% for InGaN/GaN QDs.\cite{ref27, ref28, ref29} Hence, the conclusion to be drawn from these findings is that the polarized emission from the III-nitride QDs is significantly more sensitive to asymmetry than the corresponding III-arsenide QDs. 

\begin{figure}[top]
\includegraphics[width=8 cm]{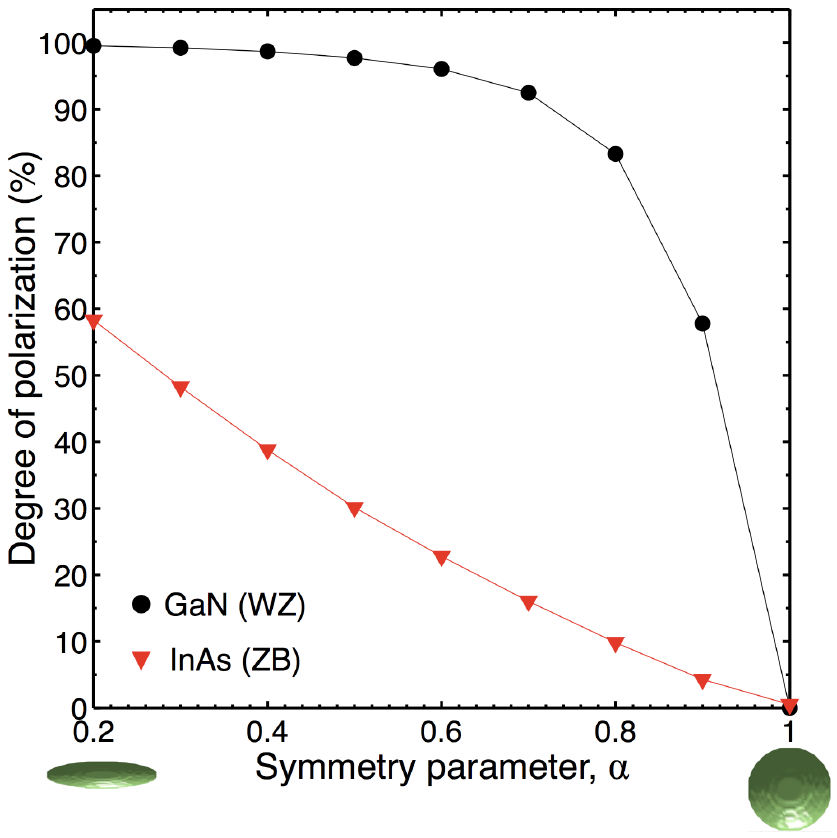}
\caption{
(Color online) The calculated degree of linear polarization for the ground state related interband transition for lens-shaped GaN/AlN and InAs/GaAs QDs as a function of the symmetry parameter, $\alpha$, while keeping the mean radius, $r$ = 6 nm constant. Solid lines serve as guide to the eye. 
}
\end{figure}

\section{The polarization dependence on an external electric field}

As discussed in the previous sections, the nitride materials are more advantageous than the arsenide materials for obtaining a high polarization degree with weakly asymmetric QDs. It is interesting to investigate if an external electric field of realist strength can provide sufficient asymmetry for III-nitride QDs in order to induce and/or change the linear polarization. This would be a desired property since it would enable dynamic control of the polarization simply by electric fields.\cite{ref06} For comparison, both wurtzite (on (0001) plane) and zinc-blend (on (111) plane) GaN and InN QDs will be studied. It should be noted that the zinc-blende QD Hamiltonian possesses three symmetry planes (point group $C_{3v}$), while the wurtzite QD Hamiltonian has rotational symmetry. However, the computed probability density functions of the hole ground state, $| \varphi |^2$ exhibit circular shapes for both the wurtzite and zinc-blend GaN QD structures (see the top panel of Fig. 3(a) and 3(b)), indicating that the hole confinement potential approximately has rotational symmetry also for zinc-blende structure when no external electric field is applied. Fig. 4 shows that the computed polarization degree for symmetric GaN QDs indeed is tunable in the range 0- $\sim$10\% by a lateral electric field up to 200 kV/cm applied in the $y$-direction, irrespective of crystal structure (see the bottom inset of Fig. 4). The confinement potential becomes weakly anisotropic under applied electric field, as probed by a small elongation of the hole probability density functions $| \varphi |^2$, in addition to the expected lateral translation of the hole center of mass, as shown in the middle panel of the Fig. 3(a) and 3(b). The dominant polarization direction correlates with the direction of the $| \varphi |^2$ elongation as shown in the bottom panel of the Fig. 3(a) and 3(b). 

A very interesting consequence of the three-fold symmetry of the zinc-blende QD is that the degree of polarization switches signs when the direction of the electric field is reversed (see Fig. 4). Thus, the dominant polarization direction can be changed by 90 degrees without the need to rotate the orientation of the electric field. On the contrary, the sign of the polarization degree remains unchanged upon reversal of the electric field for the more symmetric wurtzite structure. 

The electric field has a significantly stronger effect on the polarization of the InN QDs, as shown for the wurtzite structure in the upper inset of Fig. 4 as well as in Fig. 3(c), enabling tunability of the polarization degree between 0 - $\sim$80\% by an electric field up to 200 kV/cm. This implies that strongly linearly polarized transitions can be induced by the electric field. The higher sensitivity of the polarization on the electric field for InN is attributed to the smaller value of $\Delta_{SO}$ = 5 meV, to be compared with $\Delta_{SO}$ = 17 meV for GaN.

\begin{figure}[top]
\includegraphics[width=8.5 cm]{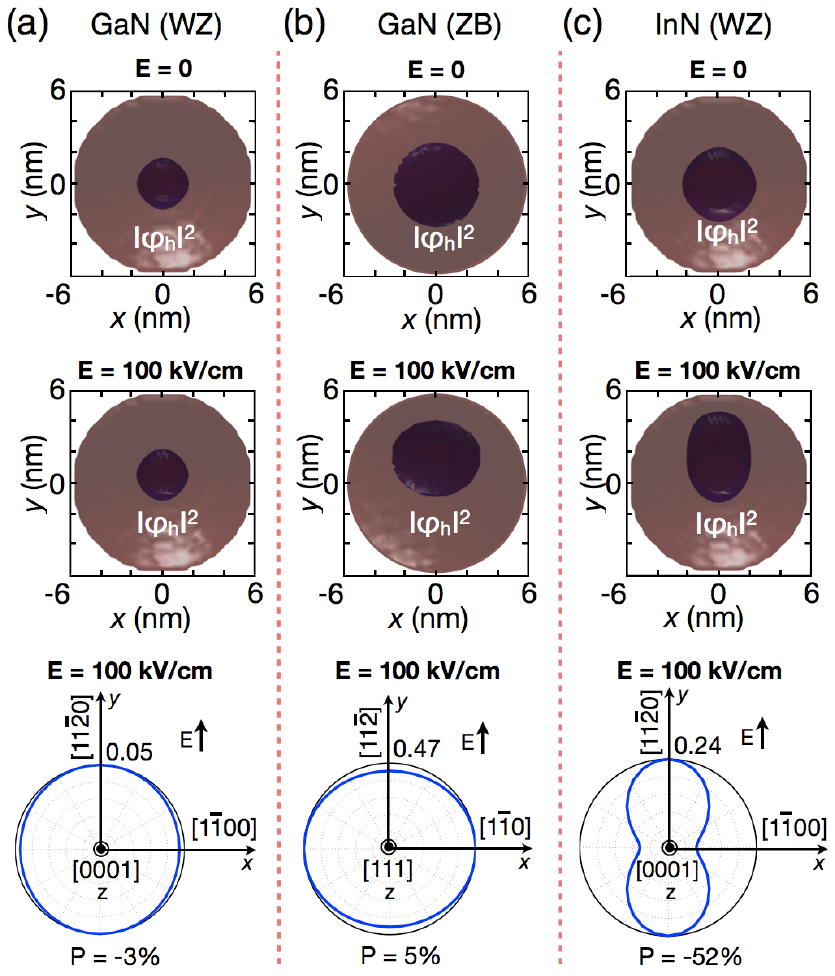}
\caption{
(Color online) The top panel illustrates the probability density function of the hole ground state $| \varphi |^2$, represented by dark isosurfaces corresponding to 20\% of the maximum value of the $| \varphi |^2$, without an external electric field applied across the QDs of three material systems, (a) wurtzite GaN (0001); (b) zince-blende GaN (111); and (c) wurtzite InN (0001). The middle panel shows the same as the top panel, but for an applied external electric field of 100 kV/cm; the bottom panel is a polar plot of the linear polarization dependence of the ground state related transition under applied electric field. The coordinates $x$, $y$, and $z$ refer to the $[1\bar{1}00]$, $[11\bar{2}0]$, and $[0001]$  ($[1\bar{1}0]$, $[110]$, and $[111]$) crystallographic directions in a wurtzite (zinc-blende) structure, respectively. 
}
\end{figure}

\begin{figure}[top]
\includegraphics[width=8 cm]{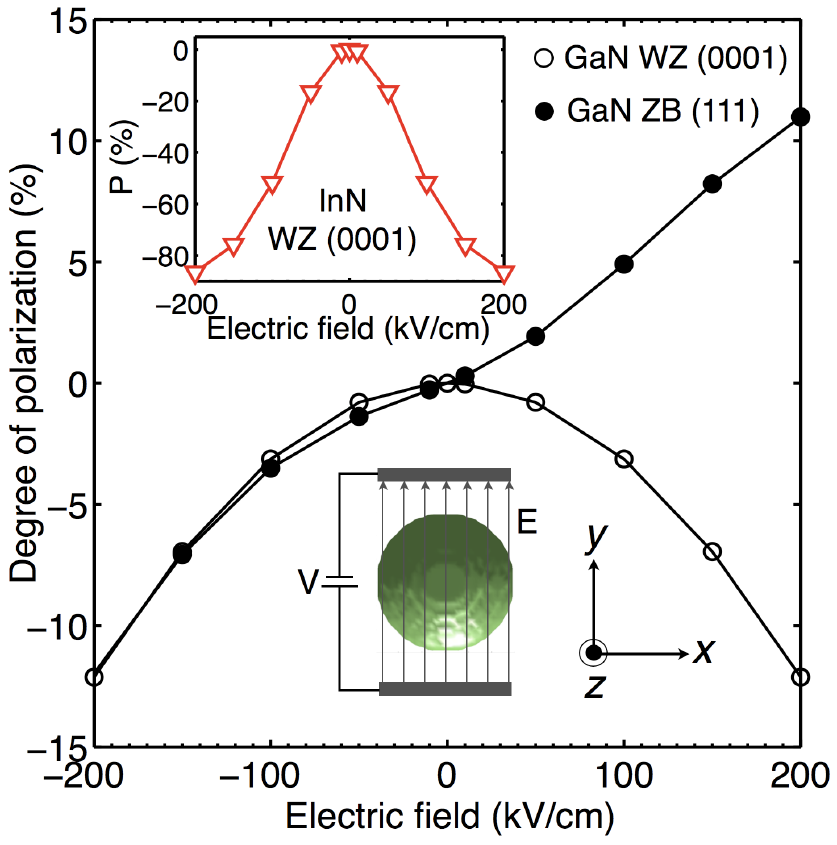}
\caption{
(Color online) The calculated degree of linear polarization for the ground state related interband transition for symmetric lens-shaped wurtzite GaN QDs on the (0001) plane and zinc-blende GaN QDs on the (111) plane as a function of an externally applied electric field ($E$). The top inset shows the same polarization dependence as in the main figure but for wurtzite InN QDs on the (0001) plane. The bottom inset shows a schematic illustration of an applied external electric field across the QD. The coordinates $x$, $y$, and $z$ refer to the $[1\bar{1}00]$, $[11\bar{2}0]$, and $[0001]$  ($[1\bar{1}0]$, $[11\bar{2} ]$, and $[111]$) crystallographic directions in a wurtzite (zinc-blende) structure, respectively. 
}
\end{figure}

Surprisingly, the degree of polarization of the zinc-blende InN QDs (symmetric QDs, $\alpha$ = 1) is dramatically more sensitive to an electric field than the corresponding wurtzite InN QDs, and a polarization degree of 98\% can be achieved with an field strength of merely 20 kV/cm (see Fig. 5). Like the GaN zinc-blende QDs, the degree of polarization of the InN zinc-blende QDs switches sign for reversed electric field. The reason for the extremely high sensitivity of the polarization upon symmetry breaking of the InN zinc-blende QDs is related to a strong in-plane anisotropy experienced by the confined hole, as revealed by the probability density function $| \varphi |^2$ in Fig. 6(a). Thus, unlike the GaN QDs, the three-fold symmetry of the zinc-blende structure is very pronounced for the hole states in InN QDs with $| \varphi |^2$ distributed equally over three arms. This leads to a radical modification of the confinement potential upon application of an electric field, which decreases the potential for one or two of the arms and accordingly $| \varphi |^2$  becomes unevenly localized mainly in one or two of the three arms (see the hole probability density functions in Fig. 6(b-c)). The very pronounced tripartite potential of InN/GaN QDs is attributed to the large strain (the lattice mismatch is 10\% for InN/GaN, to be compared with 1.4\% for GaN/AlN), which, via the deformation potential and piezoelectric fields, modifies the potential accordingly to the $C_{3v}$ symmetry.

QDs of perfectly symmetric shape are not expected in real samples, instead some degree of symmetry breaking is likely to occur. Fig. 5 shows the polarization dependence on an external electric field also for weakly asymmetric InN QDs ($\alpha$ = 0.95, 0.90), with the lateral field applied both in the $x$- and $y$-directions. These results demonstrate that the polarization resulting from a structural elongation of the QD can be compensated by an external electric field, restoring the unpolarized condition at a sufficiently high bias ($<$100 kV/cm for the QD models used here). Thus, electric field induced polarization switching is predicted to be possible also for real weakly asymmetric InN QDs.

\begin{figure}[top]
\includegraphics[width=8 cm]{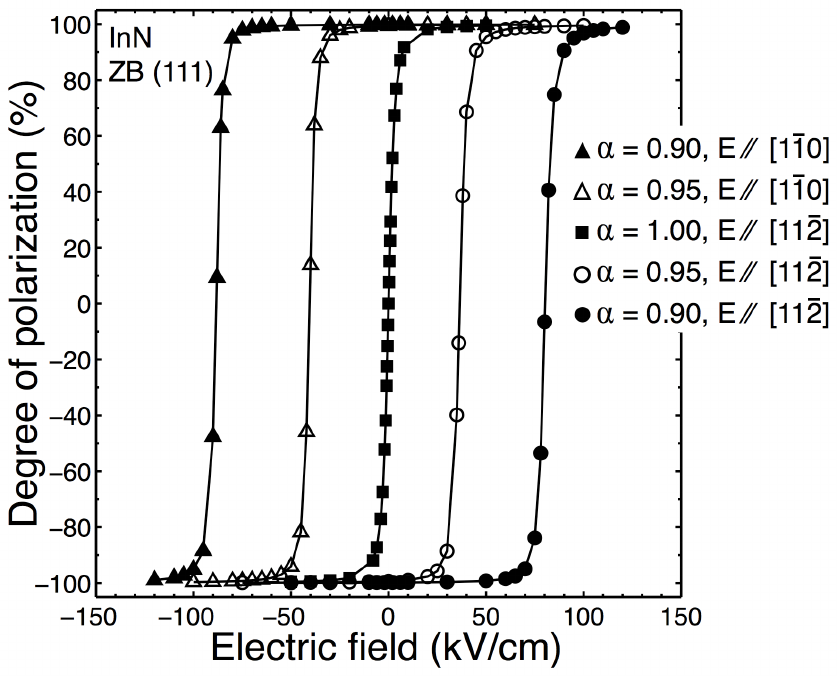}
\caption{
(Color online) The calculated degree of linear polarization of the ground state related interband transition for lens-shaped zinc-blende InN QDs on the (111) plane for different values on the symmetry parameter, $\alpha$ = 0.90, 0.95, and 1.00, as a function of an externally applied electric field in two different directions, the $[1\bar{1}0]$, $[11\bar{2}]$ directions. Solid lines serve as guide to the eye. 
}
\end{figure}

\begin{figure}[top]
\includegraphics[width=6.5 cm]{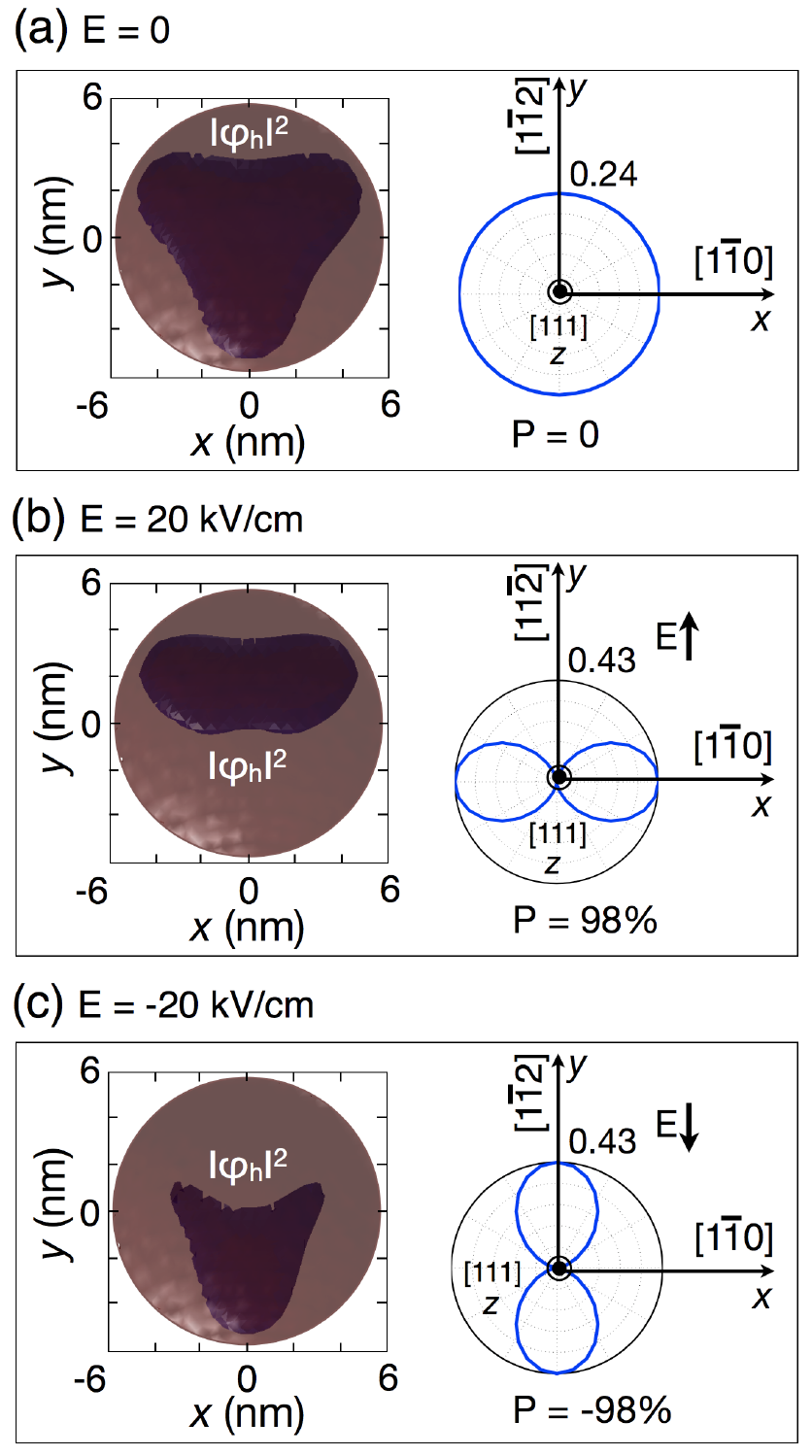}
\caption{
(Color online) (a) The probability density function of the hole ground state $| \varphi |^2$, represented by dark isosurfaces corresponding to 20\% of the maximum value of the $| \varphi |^2$, showing a triangual-like shape for the case of no electric field applied (right figure) and a polar plot of the linear polarization dependence for the ground state related interband transition (left figure) for lens-shaped zinc-blende InN on the (111) plane. The coordinates $x$, $y$, and $z$ refer to the $[1\bar{1}0]$, $[11\bar{2}]$ and $[111]$ crystallographic directions, respectively. (b) and (c) idem as (a) but for an applied external electric field of 20 kV/cm and -20 kV/cm, respectively.
}
\end{figure}

The computed ground state related transition energy for the InN QDs modeled here is 1.218 eV, and the corresponding Stark shift induced by the lateral electric field is below 10 meV for fields up to 50 kV/cm. Moreover, the energy spacing between the ground and excited hole states depends almost linearly on the field strength; 13 meV ($\sim$5.5 meV) for 50 kV/cm (20 kV/cm), indicating that cryogenic temperatures are required in order to avoid thermal population of excited hole states. As the electric field tunes the nominally unpolarized QD emission to be polarized, the intensity for the dominant polarization direction increases to almost the double (see the polar plots in Fig. 6), but the total angle integrated intensity tends to slightly decrease due to a reduced overlap between the electron and hole wavefunctions in the presence of the field. The fact that a nearly perfect linear polarization can be induced by moderate external electric fields, and its direction can be switched by 90 degrees simply by reversing the direction of the field, without the need for additional electrodes, makes the zinc-blende InN QDs potentially suitable as single photon emitters with direct polarization control on a photon-by-photon level. Such light sources are of great interest for physics experiments as well as for implementing various protocols of quantum key distribution, where the individual photons are encoded by polarization.\cite{ref01, ref30}

It should be noted that the material parameters of zinc-blende InN currently are very uncertain and their values are mainly based on theoretical estimates \cite{ref19,ref21}. Therefore, the validity of the presented prediction can be questioned. However, it is a very general result that for a confining potential resulting in $| \varphi |^2$ distributed equally along three arms (or three short quantum wires), the hole localization will be very sensitive to perturbations induced by e.g. an electric field. For a material with a small value for $\Delta_{SO}$, such a perturbation leads to almost perfect linear polarization of the optical interband transitions. Thus, efficient polarization switching is predicted also for GaN QDs of triangular shape. We hope that this theoretical work will inspire future experiments on the optical polarization properties of III-nitride QDs.

\section{Conclusions}

In conclusion, the split-off energy has been highlighted as the main material parameter influencing the optical linear-polarization properties of asymmetric QDs. Therefore, III-nitrides (e.g. GaN, InN and AlN) based QDs are more promising as sources of linearly polarized light than the conventional III-arsenides (e.g. GaAs, InAs and AlAs) based QDs.  Dynamic control of the linear polarization by a variable lateral electric field was proposed, and InN was demonstrated to have the best prospects for this purpose. The zinc-blende InN QD is particularly advantageous, since the symmetry of the zinc-blende crystal enables switching of the polarization by 90 degrees by simply reversing the electric field, without the need to change its orientation. Moreover, the particular confinement potential created within a lens-shaped InN QD will make it extremely sensitive to an applied electric field, with a sensitivity about 50 times higher than for a corresponding GaN QD. Thus, InN QDs are proposed to be used as single-photon emitters that allow direct control of the linear polarization on a photon-by-photon level. Such light sources are of great interest for physics experiments as well as for implementing various protocols of quantum key distribution.

\section{Acknowledgements}

This work has been supported by a Ph.D. scholarship from Thaksin University in Thailand for S. Amloy, grants from the Swedish Research Council (VR), the Nano-N consortium funded by the Swedish Foundation for Strategic Research (SSF), and the Knut and Alice Wallenberg Foundation. We acknowledge support from the Swedish Government Strategic Research Area in Materials Science on Functional Materials at Link\"oping University (Faculty Grant SFO-Mat-LiU \# 2009-00971). 

\bibliography{polarization_refs}

\begin{thebibliography}{29}
\bibitem[{not()}]{note1}
\bibinfo{note}{The values used for the valence band offsets are; 0.85 eV (0.58
  eV) between the wurtzite (zinc-blende) GaN and AlN, 0.50 eV (0.3 eV) between
  the wurtzite (zinc-blende) InN and GaN and 0.21 eV between the zinc-blende
  InAs and GaAs. The isotropic dielectric constants used are 9.8 for GaN, 13.2
  for InN, and 14.6 for InAs.}
\end{thebibliography}

\end{document}